\documentclass{article}

\usepackage[a4paper, total={6in, 8in}]{geometry}
\usepackage{graphicx}
\usepackage{subcaption}
\usepackage{hyperref}
\usepackage{color}
\usepackage{acronym}

\newcommand{\rfc}[1]{RFC #1 \cite{RFC#1}}

\newcommand{\jp}[1]{\textcolor{black}{#1}}

\graphicspath{{./figures/}}

\newcommand\fig[1]{Figure~\ref{fig:#1}}

\newcommand\sect[1]{Section~\ref{sec:#1}}

\begin{document}

\title{OIDC²: Open Identity Certification with OpenID Connect}
\author{Jonas Primbs and Michael Menth}

\maketitle

\begin{abstract}
\ac{oidc} is a widely used authentication standard for the Web.
In this work, we define a new \ac{ict} \jp{to enable \ac{e2e} user authentication by using and extending \ac{oidc}'s native mechanisms}.
An \ac{ict} can be thought of as a JSON-based, short-lived user certificate for \ac{e2e} user authentication without the need for cumbersome key management.
A user can request an \ac{ict} from his \ac{op} and use it to prove his identity to other user\jp{s t}hat trust the \ac{op}.
We call this approach \ac{oidc²} and compare it to other well-known \ac{e2e} authentication methods.
Unlike certificates, \ac{oidc²} does not require installation and can be easily used on multiple devices, making it more user-friendly.
We outline protocols for implementing \ac{oidc²} based on existing standards.
We discuss the trust relationship between entities involved in \ac{oidc²}, propose a classification of \acp{op}' trust level, and propose authentication with multiple \acp{ict} from different \acp{op}.
We explain how different applications such as video conferencing, \jp{\ac{im}}, and email can benefit from \acp{ict} for \ac{e2e} authentication and recommend validity periods for \acp{ict}.
To test \ac{oidc²}, we provide a simple extension to existing \ac{oidc} server software and evaluate its performance.

\end{abstract}

%-------------------------------------------------------------------------------
% \vspace{-0.2cm}
\section{Introduction}
\label{sec:introduction}
% \vspace{-0.2cm}
%-------------------------------------------------------------------------------
% Einführung, warum E2E Authentication notwendig ist:
In most communication services, users identify each other through account profiles in which they provide their own identity information.
To make these profiles more trustworthy, social network operators such as Meta and \jp{X} offer identity verification services for an additional fee that can only be used within their ecosystem.
However, identity verification is often a cumbersome process that users may not want to repeat for each of their service platforms.
\jp{Furthermore}, users must still trust the service provider to sufficiently verify identities and not impersonate them.
\ac{e2e} user authentication mechanisms \jp{like X.509 certificates or \ac{pgp}} attempt to solve this problem, but often lack adoption due to poor usability \jp{caused by the necessity of manual key management}.
% Vorstellung OIDC und SSO; Hinweis darauf, dass E2E Authentication fehlt:
Therefore, reusing an account \jp{with} a verified identity \jp{for \ac{e2e} user authentication} would be desirable.
With modern \ac{sso} services, users can reuse their existing accounts to log in to other services.
The \ac{oidc} protocol, which is based on the OAuth 2.0 authorization framework, is widely used for this purpose.
However, \ac{oidc} is designed for user-to-service authentication and does not address the purpose of \ac{e2e} user authentication.

% Vorstellung ICT und OIDC²:
\jp{In this paper, we developed an \ac{oidc} extension for \ac{e2e} user authentication which is (1) easy to implement and (2) easy to use.}
\jp{Therefore}, we define a new \ac{ict} for \ac{oidc}.
It is similar to the \ac{idt} which holds identity claims about a user, but also contains a verified public key of the user's client.
As such, it can be thought of as a JSON-based, short-lived user certificate without the need for a revocation mechanism.
The use of an \ac{ict} differs significantly from the use of an \ac{idt}.
A user requests an \ac{ict} from his \ac{op} and presents it to another user's client to authenticate himself.
If the other user trusts the issuing \ac{op}, his client verifies the integrity and validity of the \ac{ict} and authenticates the user using his client's public key contained in the \ac{ict}.
As the \ac{op} certifies the identity of the user, we call this concept \acp{oidc²}.
It facilitates mutual authentication of users if they trust each other's \ac{op}.

% Notwendigkeit von Identitätsverifizierung und Relevanz von OIDC:
\jp{Although} most \acp{op} have a rather superficial identity verification process for their accounts, some practice a more thorough verification.
In particular, new players such as banks and government institutions that perform rigorous identity verification for their accounts are becoming \acp{op}.
With \ac{oidc²}, unknown users can be reliably authenticated if they have an \ac{oidc} account at a trusted \ac{op}.
% Kritik an PKI und SSI:
Some services already provide strong user authentication, but these methods are difficult to use.
Many \ac{im} services support the exchange of public keys between users when they meet in person.
\acp{pki} require certificate management by users and reliable revocation list checking.
\ac{pgp} or \ac{smime} have long been proposed for email authentication, but are rarely used \cite{Stransky2022}.
\ac{ssi} technology is currently emerging and solves this problem with device-specific long-term keys in a wallet app.
However, this requires not only revocation mechanisms, but also recovery mechanisms in case the phone with the wallet app is lost or stolen.
% Was OIDC² besser macht und wie:
\ac{oidc²} provides a more user-friendly alternative for \ac{e2e} authentication.
The \ac{ict} is short-lived, eliminating the need for cumbersome key revocation mechanisms, which improves security.
\ac{oidc²} avoids complex key management across devices by simply requesting a new \ac{ict} from the \ac{op} whenever needed.
Using trusted \acp{op} that verify the identity of their users also eliminates the need for face-to-face key exchange.
Thus, a trusted \ac{op} can be compared with a trusted certification authority in a \ac{pki} or a trusted issuer in the \ac{ssi} context.
However, \ac{oidc²} is only a lightweight extension for \ac{e2e} authentication with existing \ac{oidc} accounts.
It is not intended to replace \jp{X.509} \acp{pki} or \acp{ssi}.

% Vorstellung der Paper-Struktur:
The paper is structured as follows.
In \sect{background}, we revisit \jp{the} basics of OAuth 2.0 and \ac{oidc}, and in \sect{related_technologies}, we review related authentication technologies.
\sect{oidc2} introduces the concept of \ac{oidc²} and proposes the extension to the \ac{oidc} protocol.
Trust relationships in \ac{oidc²}, a classification of \acp{op}, authentication with multiple \acp{ict}, and validity periods of \acp{ict} are discussed in \sect{security}.
In \sect{applications}, we explain how \ac{oidc²} can be applied to video conferencing, \ac{im}, and email.
To test \ac{oidc²}, we provide a simple extension to the \ac{oidc} server software in \sect{implementation}, which we evaluate in \sect{evaluation}.
\sect{conclusion} concludes our findings.

%-------------------------------------------------------------------------------
% \vspace{-0.2cm}
\section{Introduction to OAuth 2.0 and \ac{oidc}}
\label{sec:background}
% \vspace{-0.2cm}
%-------------------------------------------------------------------------------
% Überblick über Struktur der Section:
We introduce basics of OAuth 2.0 and \ac{oidc}, as they are the underlying technologies for \ac{oidc²}.
We discuss their trust relationship and explain how they facilitate \ac{sso}.
In the following, we capitalize technical terms from the official OAuth 2.0 and \ac{oidc} terminology.
For ease of understanding, we omit non-essential steps in the protocols and refer to the authoritative standards for details.

% \vspace{-0.2cm}
\subsection{OAuth 2.0}
\label{sec:oauth2}
% \vspace{-0.1cm}
%-----------------------------------
% Einführung OAuth; Begriffsdefinitionen:
The OAuth 2.0 authorization framework, as defined in \rfc{6749}, is based on \ac{http} (\rfc{7231}) and \jp{JSON (\rfc{8259})}.
It allows a user to grant his Client scoped access to his resources on a server, e.g., to only read emails.
A Client can be a web application, or a native email client application.
In OAuth 2.0, this server is called \ac{rs} because it protects the user's \acp{pr}; the user is called the \ac{ro}.

% Vorteil von OAuth gegenüber Username/Password Autorisierung; Einführung Access Token:
Without OAuth 2.0, the \ac{ro} would leave his credentials with his Client to log in directly to the \ac{rs}.
With OAuth 2.0, the \ac{ro} logs in to an \ac{as} and tells the \ac{as} to authorize the Client to access a scoped set of \acp{pr}.
To do this, the \ac{as} issues an \ac{at} to the Client.
This \ac{at} allows the Client to access the \acp{pr} on the \ac{rs}.
In this way, OAuth 2.0 improves the security by granting Clients only a scoped access to the user's account without exposing the user's credentials to any of the Clients.

\begin{figure}[ht]
    % \vspace{-0.5cm}
    \centering
    \begin{subfigure}[b]{.54\linewidth}
        \centering
        \includegraphics[width=\linewidth]{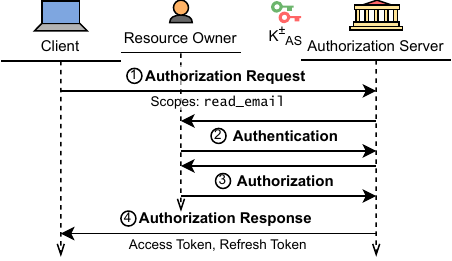}
        \caption{Simplified Authorization Request.}
        \label{fig:oauth_authorization_flow}
    \end{subfigure}
    \hfill
    \begin{subfigure}[b]{.44\linewidth}
        \centering
        \includegraphics[width=\textwidth]{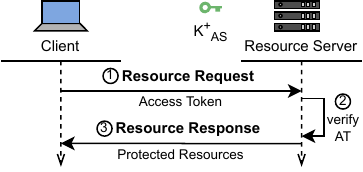}
        % \vspace{0.9cm}
        \caption{Resource Request.}
        \label{fig:oauth_resource_request}
    \end{subfigure}
    \caption{OAuth 2.0 protocol flows.}
    % \vspace{-0.6cm}
\end{figure}

% Wie das Access Token abgerufen wird:
\fig{oauth_authorization_flow} shows a simplified authorization request where the \ac{ro} authorizes his Client to read email.
First, the Client requests access to the Scope \texttt{read\_email}, which authorizes read-only access to the \ac{ro}'s emails (1).
Then, the \ac{as} authenticates the \ac{ro} (2) and the \ac{ro} authorizes the Client for the requested Scope (3).
Finally, the \ac{as} issues the \ac{at} and optionally a \ac{rt} (4).
This \ac{at} contains the authorized Scopes with a short validity period.
It is signed with the \ac{as}'s private key $K^-_{\ac{as}}$.
The \ac{rt} is like a revocable session ID that authorizes the Client to refresh an expired \ac{at} without further user interaction.

% Wie das Access Token einegsetzt wird um Protected Resources abzurufen:
\fig{oauth_resource_request} describes a Resource Request where the Client uses the \ac{at} to access \acp{pr} on the \ac{rs}.
First, the Client requests the \acp{pr} and provides the \ac{at} to prove authorization (1).
Then, the \ac{rs} verifies the \ac{at} for a sufficient Scope, its expiration date, and the validity of its signature with the \ac{as}'s public key $K^+_{\ac{as}}$ (2).
Finally, the \ac{rs} responds with the \acp{pr} (3).

% \vspace{-0.2cm}
\subsection{OpenID Connect (OIDC)}
\label{sec:oidc}
% \vspace{-0.1cm}
%-----------------------------------
% Einführung OpenID Connect; Abgrenzung zu OAuth:
\ac{oidc} \cite{OidcCore} is an authentication framework that allows users to be authenticated with an \ac{sso} identity through a third-party service, such as an email service.
It extends OAuth 2.0 for this purpose.
Unlike the example in \sect{oauth2}, the \ac{sso} identity has no relationship to the third-party service.

% \vspace{-0.8cm}
\begin{figure}[ht]
    \centering
    \begin{subfigure}[b]{.50\linewidth}
        \centering
        \includegraphics[width=\linewidth]{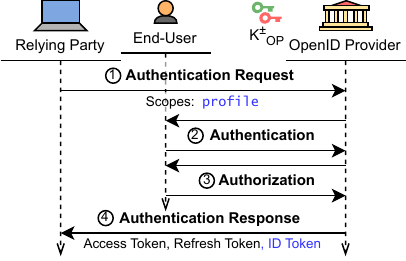}
        \caption{Simplified Authentication Request.}
        \label{fig:oidc_authentication_flow}
    \end{subfigure}
    \hfill
    \begin{subfigure}[b]{.49\linewidth}
        \centering
        \includegraphics[width=\linewidth]{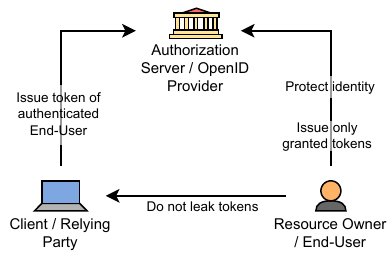}
        \caption{Trust relationship among involved entities in OAuth 2.0 and OIDC.}
        \label{fig:oidc_trust_relationship}
    \end{subfigure}
    \caption{OpenID Connect Authentication Flow and trust relationship.}
    % \vspace{-0.6cm}
\end{figure}

% Begriffsdefinitionen mit Abgrenzung zu OAuth:
In \ac{oidc}, an \ac{eu} is authenticated by an \ac{op}.
The \ac{eu} grants \ac{oidc}-specific Scopes to the EU's intended service, e.g., to his email client, which is called the \ac{rp}.
This communication flow is supported by OAuth 2.0, where the \ac{eu} corresponds to a \ac{ro}, the \ac{op} to an \ac{as}, and the \ac{rp} to a Client.
The \ac{op} issues an \ac{idt} to the \ac{rp}.
This \ac{idt} contains claims about the authentication event which typically includes information about the \ac{eu}, such as his name or address.
Since this Authorization Request is for authentication, it is called an Authentication Request in \ac{oidc}.
With this mechanism, an \ac{eu} can be authenticated with his \ac{sso} identity by different services without providing his credentials.
Instead, the \ac{op} passes profile information about the \ac{eu} to the \ac{rp} as identity claims in the \ac{idt}, but the \ac{eu} controls which information is passed.

% Wie das ID Token abgerufen wird:
\fig{oidc_authentication_flow} describes a simplified Authentication Request where the \ac{eu} is authenticated by the \ac{rp} via the \ac{op}.
First, the \ac{rp} requests access to the \texttt{profile} Scope.
If the \ac{eu} grants this Scope, the \ac{rp} is authorized to access the \ac{eu}'s profile information (1).
Then, the \ac{eu} is authenticated by the \ac{op} (2) and authorizes the \ac{rp} for the requested Scope (3).
Finally, the \ac{op} issues an \ac{idt} in addition to the \ac{at} and an optional \ac{rt} (4).
This \ac{idt} contains the identity claims related to the authorized \texttt{profile} Scope, such as the \ac{eu}'s name, profile picture, or date of birth, and is signed with the \ac{op}'s private key $K^-_{\ac{op}}$.
The \ac{rp} can verify the signature of the identity claims with the \ac{op}'s public key $K^+_{\ac{op}}$.

% \vspace{-0.2cm}
\subsection{Trust Relationship}
\label{sec:trust_relationship}
% \vspace{-0.1cm}
%-----------------------------------
% Einführung in Trust-Relationship; Referenz auf Abbildung:
The following describes the resulting trust relationship between entities in OAuth 2.0 and \ac{oidc} as shown in \fig{oidc_trust_relationship}.

% Trust-Verhältnis Client/RP <-> AS/OP:
The Client/\ac{rp} never sees any credentials from the \ac{ro}/\ac{eu} because the authentication process is performed solely by the \ac{as}/\ac{op}.
Therefore, the Client/\ac{rp} must rely on the \ac{as}/\ac{op} to correctly verify the identity of the \ac{ro}/\ac{eu} and that the \ac{as}/\ac{op} will issue the correct \ac{at}/\ac{idt} of the authenticated \ac{ro}/\ac{eu}.
% Trust-Verhältnis RO/EU <-> Client/RP:
Once the Client/\ac{rp} of the \ac{ro}/\ac{eu} receives the tokens, the \ac{ro}/\ac{eu} may not be able to verify what it is doing with them.
The Client/\ac{rp} may even leak the tokens, so the \ac{ro}/\ac{eu} must trust that it is working as intended.
To minimize this risk, the \ac{ro}/\ac{eu} restricts the Client/\ac{rp}'s access to only the necessary \acp{pr} and identity claims.
% Trust-Verhältnis AS/OP <-> RO/EU:
The \ac{ro}/\ac{eu} must also trust the \ac{as}/\ac{op} to protect his identity.
This includes a secure login process and secure credential storage, but also that the \ac{as}/\ac{op} will not impersonate his account.
Such impersonation would not even require any credentials since the \ac{as}/\ac{op} needs only its private key $K^-_{\ac{as}}$/$K^-_{\ac{op}}$ to sign an \ac{at}/\ac{idt}.

% \vspace{-0.2cm}
\subsection{Single Sign-On with OAuth 2.0 and OIDC}
\label{sec:sso}
\begin{figure}[ht]
    \centering
    \includegraphics[width=\linewidth]{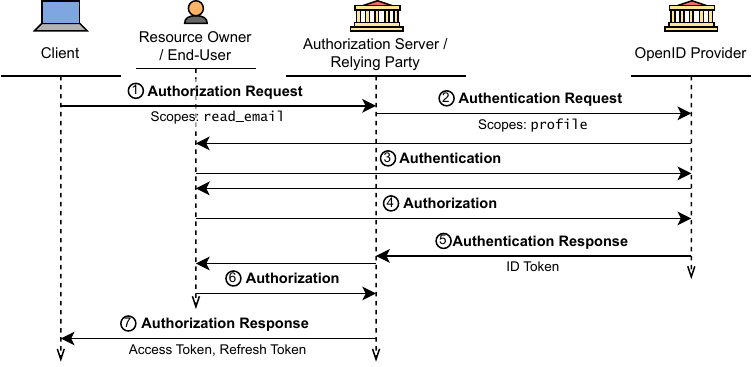}
    \caption{Simplified authentication to an \ac{as} with \ac{oidc} and authorization of a Client with OAuth 2.0.}
    \label{fig:oidc_sso}
    % \vspace{-0.4cm}
\end{figure}

% \vspace{-0.1cm}
%-----------------------------------
% Einführung in Single Sign-On; Referenz auf Abbildung:
Today, many services require dedicated accounts, forcing users to remember multiple service-specific credentials.
With \ac{sso} systems, users only need to remember the credentials for one account.
They can use this \ac{sso} identity to log in to multiple service accounts.
Logging in to a service account with this \ac{sso} identity is typically solved with a combination of OAuth 2.0 and \ac{oidc}, as depicted simplified in \fig{oidc_sso}.

% Beschreibung des OAuth Ablaufs in Abbildung
First, the Client initiates an OAuth Authorization Request to the service-specific \ac{as} (1).
Instead of using service account credentials, the \ac{ro} chooses to log in with his \ac{sso} identity via \ac{oidc}.
To do this, the \ac{as} acts as a \ac{rp} and initiates an \ac{oidc} Authentication Request to the \ac{op} (2).
The \ac{eu} is then authenticated by the \ac{op} with his credentials (3) and consents to the \ac{op} providing the \ac{rp} with access to his profile information (4).
Technically, this consent is an authorization in the sense of OAuth 2.0.
The \ac{op} then responds with an \ac{idt} to the \ac{rp} (5), which authenticates the \ac{eu} to the \ac{as} and completes the \ac{oidc}-based authentication process.
Now the authenticated \ac{ro} authorizes the requested Scopes of the Client (6).
Finally, the \ac{as} issues an \ac{at} and an optional \ac{rt} to the Client (7).

% Benutzer muss jetzt auch OP vertrauen, welcher dadurch mächtiger wird als AS:
In an \ac{sso} environment, the trust relationship changes slightly.
While the user has to trust the \ac{as} not to impersonate his service account, he also has to trust the \ac{op} not to impersonate any of his service accounts.
This makes the \ac{op} very powerful because it could impersonate any of the user's service accounts.
Therefore, \acp{eu} should only choose trusted \acp{op}.

%-------------------------------------------------------------------------------
\section{Related Technologies}
\label{sec:related_technologies}
%-------------------------------------------------------------------------------
% Überblick über diese Section:
We review related technologies for \ac{e2e} authentication and compare them to \ac{oidc²}.

\subsection{Identity Providers and Certificates}
\label{sec:idp}
%-----------------------------------
% Einführung PKI, X.509, S/MIME und Nachteile:
In a \ac{pki} \cite{Weise2001}, a \ac{ca} verifies that an entity's real-world identity and long-term public key $K^+_E$ belong together, records them in a document, signs it, and issues it in the common X.509 certificate format (\rfc{5280}).
Such X.509 certificates are used, e.g.\jp{,} in the \ac{smime} standard (\rfc{8551}) to authenticate and encrypt email.
However, \jp{the identity verification is cumbersome and users have to manage their own certificate files, an authentication concept that many people are still unfamiliar with, which is why} only $2.50 \%$ of over $81$ million emails examined in a study \cite{Stransky2022} were signed with \ac{smime}.

% Einführung SAML2-basierter Cisco Draft:
To simplify this process, Cisco proposed an expired Internet draft \cite{I-D.biggs-acme-sso} where an \ac{idp} issues X.509 certificates to its users.
According to their white paper \cite{WebexWhitepaper}, \jp{these certificates are used for challenge/response-based \cite{Kushwaha2021}} \ac{e2e} user authentication in the Webex video conferencing service \jp{where session partners trust each other's \ac{op}}.
The draft \cite{I-D.biggs-acme-sso} is designed for the \ac{saml2} authentication standard \cite{saml2core}, but \ac{oidc} performs better for mobile devices and cloud computing \cite{Naik2016}.
This may be one reason why the design has not been adopted by other applications and \acp{idp}.
% Abgrenzung von OIDC² zu Cisco draft:
Conceptually, the presented approach is similar to \ac{oidc²}; we continue with the differences.
X.509 is a binary format limited to a small set of standardized identity-related fields \cite{RFC8551}\jp{, while \ac{oidc²} uses the more flexible JSON-based \ac{jwt} (\rfc{7519}) format} to represent claims about the user.
\jp{These claims are signed by the \ac{idp} and their formats are standardized in the \ac{oidc} Core \cite{OidcCore} and eKYC \cite{OidcEkyc} specifications.}
\jp{In addition, long-lived user certificates require key revocation mechanisms and error-prone manual key management by the user, whereas short-lived \acp{ict} are issued spontaneously and only require the user to log in.}
\jp{This reduces security issues and improves the user experience}.

\subsection{Self-Sovereign Identity (SSI)}
\label{sec:ssi}
%-----------------------------------
% Einführung in SSI; Erklärung der Verwendung und Speicherung von public/private key:
In \ac{ssi} \cite{Muehle2018}, participating entities generate their own asymmetric key pairs $K^\pm$.
Entities are identified by their \ac{did}, which is linked to at least one public key $K^+$.
Entities store their private key $K^-$ in their digital wallet, e.g.\jp{,} an app on their smartphone.
This can be used for \ac{e2e} authentication with the key pair $K^\pm$.

% Begriffsdefinition Holder, Issuer, Verifier, Verifiable Credential, Verifiable Presentation:
\ac{ssi} describes three entities: the \ac{issuer}, the \ac{holder}, and the \ac{verifier}.
The issuer knows or verifies the \jp{holder's credentials} and issues them to the holder as a \ac{vc}.
This \ac{vc} is signed by the issuer with his private key $K^-_{\ac{issuer}}$; it contains the issuer's ${DID}_{\ac{issuer}}$ and the credentials and ${DID}_{\ac{holder}}$ of the holder.
The holder holds this \ac{vc} in his wallet and presents it to a verifier as a \ac{vp}.
This \ac{vp} is signed by the holder with his private key $K^-_{\ac{holder}}$; it contains the \ac{vc} and the verifier's ${DID}_{\ac{verifier}}$.
The verifier verifies this \ac{vp} by checking the issuer's signature on the \ac{vp} and the issuer's signature on the \ac{vc}.
If the verifier accepts the issuer as a trusted authority for issuing the holder's credentials, then the verifier trusts that these credentials belong to the holder.

% Implementierungen mit Blockchain und OIDC; Beispiel US Mobile Driving License:
Early implementations of \ac{ssi} made use of blockchain technology \cite{Ferdous2019} and used a \jp{distributed public} ledger \cite{Ioini2018} to store the mapping of a {\ac{did}} to its associated public keys.
Modern approaches are based on OAuth 2.0 and \ac{oidc}, such as the mobile driving license in the United States standardized in ISO/IEC 18013-5:2021 \cite{ISO18013-5}.
This approach implements the \ac{siopv2} \cite{siopv2} draft in the wallet app for key management.
Driving license offices provide OAuth 2.0 based interfaces defined in the \ac{openid4vci} draft \cite{openid4vci} to issue driving licenses as \acp{vc} in the \ac{w3c} format \cite{w3cVc}.
Drivers present these \acp{vc} as \acp{vp} to police officers using OAuth 2.0 based interfaces between smartphones defined in the \ac{openid4vp} draft \cite{openid4vp}.
Another \ac{oidc} draft describes the issuance of identity claims of the \ac{idt} as a \ac{vc} \cite{userinfoVc}.
This is similar to our approach, but requires the full \ac{openid4vci} infrastructure to be deployed, which is currently rare.

% Probleme an SSI Lösungen: Long-term keys, die revokierbar sein müssen oder nicht übertragen werden können.
\jp{Although \ac{ssi} is now being adopted for some government use cases, there are still issues with usability \cite{Sartor2022}\cite{Zaeem2021} and identity recovery \cite{Zhou2019}.}
\jp{These stem from manual key management by users who are unaware of their responsibilities, and the entirely new concept of operation.}
Since the private key is a long-term key that could be leaked during its lifetime, the system requires a key revocation list.
But as argued by Ronald L. Rivest more \jp{than} two decades ago \cite{Rivest1998}, revocation lists should be avoided for security reasons.
Modern technologies such as \ac{hsm} or \ac{tpm} address this problem by protecting the private key inside the hardware.
Here, the private key cannot be exported and can only be used for signing after the platform integrity has been verified and the user has been authenticated.
This creates problems when a user wants to use \acp{vc} from other devices.
\jp{Additionally}, if the device is lost or broken, the user needs a recovery method for the private key and \ac{did} that must be configured in advance.

% Was OIDC² besser macht:
\ac{oidc²} does not have these problems.
It uses short-lived ephemeral key pairs and \acp{ict} \jp{that do not require a} specific hardware or software platform.
\jp{It also leverages existing account recovery capabilities and the familiar sign-in user authentication concept}.
Compared to \ac{ssi} approaches, it does not require currently rarely deployed frameworks such as installed wallet apps, issued \acp{vc}, and a huge amount of implemented new standards.
Instead, \ac{oidc²} requires a small extension of \acp{op} to use existing \ac{oidc} accounts.
% All it takes is an existing OIDC-based user account and implemented OAuth 2 and OIDC features plus our tiny extension.
In contrast, the \ac{ict} may also contain claims that the issuing \ac{op} is not a trusted source of, which will be discovered in \sect{trust_relationship}.

% \vspace{-0.2cm}
\subsection{OpenPubkey}
\label{sec:openpk}
% \vspace{-0.1cm}
%-----------------------------------
% Vorstellung von OpenPubkey:
BastionZero has developed OpenPubkey \cite{openpubkey} which is very similar to \ac{oidc²}.
The \ac{rp} of an \ac{eu} can create a \ac{cic} that contains, among others, the \ac{rp}'s public key $K^+_C$.
When requesting an \ac{idt} (see \fig{oidc_authentication_flow}), the \ac{rp} can optionally provide a nonce in the Authentication Request (1), which we omitted in \sect{oidc}.
The \ac{op} will then insert this nonce into the \ac{idt} before issuing it (4).
With OpenPubkey, the \ac{rp} offers its hashed \ac{cic} as a nonce to be inserted into the \ac{idt}.
After receiving the \ac{idt}, the \ac{rp} appends the \ac{cic} and signs it with its private key $K^-_C$, resulting in a \ac{pk} Token.
\jp{The \ac{rp} can use this \ac{pk} Token to be authenticated with the  \ac{eu}'s identity.}

However, from our point of view, this approach \jp{has some security-relevant drawbacks.}
In an \ac{sso} context, the \ac{rp} is often a login service (see the \ac{as} in \fig{oidc_sso}) that the \ac{eu} usually authorizes to access his profile information.
\jp{This \ac{rp} may act malicious and} request a \ac{pk} Token with its own public key $K^+_C$ to impersonate the \ac{eu} without his knowledge.
The authors' solution to this problem is to have the authenticating user only accept \ac{e2e} authentications \jp{from trusted \acp{rp}}, identified by \jp{their} Client ID contained in the \ac{pk} Token.
First, this \jp{induces} a high burden on the user, which is unacceptable since it is difficult for the user to identify trusted \acp{rp}.
Second, the \ac{eu}'s trust in a service, such as an online store, may be sufficient to be authenticated by that store, but it may not be sufficient to allow the store to impersonate him.
Third, in open communication systems such as email, there are many clients, and it is unlikely that all of them are trusted.
This limits the use of OpenPubkey to a small set of explicitly trusted services and clients.
We believe that these three problems \jp{may cause security vulnerabilities in the future}.
In contrast, with \ac{oidc²}, the \ac{eu} does not risk being impersonated when logging in to a malicious service.

% Was OIDC² besser macht:
\ac{oidc²} solves this problem by introducing \jp{the new \ac{ict}} that can only be requested by an \ac{rp} with sufficient scope for \jp{a specific} \ac{e2e} authentication \jp{context}.
This means that an \ac{eu} can control whether to issue only an \ac{idt} or also an \ac{ict}.

%-------------------------------------------------------------------------------
\section{OIDC²: Open Identity Certification with \ac{oidc}}
\label{sec:oidc2}
% \vspace{-0.2cm}
%-------------------------------------------------------------------------------
% Überblick über Section:
This section describes the \ac{oidc²} concept in more detail and proposes a simple \ac{oidc} protocol extension to support it.

% \vspace{-0.2cm}
\subsection{Concept of \ac{oidc²}}
\label{sec:oidc2_concept}
% \vspace{-0.1cm}
%-----------------------------------
% Überblick wie wir OIDC² erklären:
We define new terminology, introduce the \acf{ict}, and explain how to use it.

% \vspace{-0.2cm}
\subsubsection{Terminology}
\label{sec:oidc2_terminology}
% Einführung Terminologie EU, Client, AU, AP, OP und Keys:
Consider a user of one application authenticating to a user of another application.
The user authenticating himself is called the \acf{eu}, his application is called the Client.
The other user is called the \acf{au}, and his application is called the \acf{ap}.
% Erklärung der Herkunft der Terminologie; Klarstellung, dass dies nicht heißt, dass OAuth und OIDC für OIDC² erforderlich sind:
We also assume that the \ac{eu} has an \ac{sso} identity provided by an \acf{op} trusted by the \ac{au}.
The terminology used for the \ac{eu}, Client, and \ac{op} is consistent with the combined OAuth 2.0 and \ac{oidc} scenario described in \sect{sso}.
However, \ac{oidc²} does not require this scenario.

% \vspace{-0.2cm}
\subsubsection{\acf{ict}}
\label{sec:ict}
% Einführung ICT und abgrenzung von ID Token:
We introduce the \ac{ict}, which addresses the \ac{e2e} authentication use case.
The \ac{ict} contains the Client's verified public key $K^+_C$, an application-specific Scope, an expiration date, and a unique identifier of the \ac{eu}'s \ac{sso} identity.
It may also contain other claims about the user which are not necessarily verified by the \ac{op}.

% \vspace{-0.2cm}
\subsection{\ac{ict} Request}
\label{sec:obtain_ict}
\begin{figure}[ht]
    \centering
    \begin{subfigure}[b]{.47\linewidth}
        \centering
        \includegraphics[width=\linewidth]{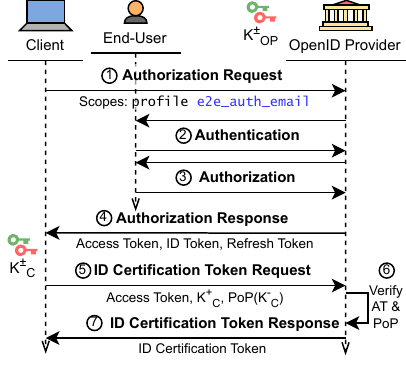}
        \caption{Simplified \ac{ict} Request.}
        \label{fig:oidc2_ict_request}
    \end{subfigure}
    \hfill
    \begin{subfigure}[b]{.52\linewidth}
        \centering
        \includegraphics[width=\textwidth]{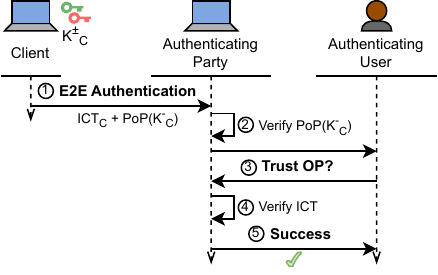}
        % \vspace{.7cm}
        \caption{\acf{e2e} authentication.}
        \label{fig:oidc2_e2ea}
    \end{subfigure}
    \caption{Protocol extension of \ac{oidc²}.}
    % \vspace{-0.6cm}
\end{figure}

% \vspace{-0.1cm}
%-----------------------------------
% Wann ein ICT abgerufen wird; Verweis auf Grafik:
The Client uniquely requests an \ac{ict} from the \ac{op} for each \ac{e2e} authentication process.
\fig{oidc2_ict_request} simplifies the \ac{ict} request.

% Abruf von Access Token:
First, the Client performs an OAuth 2.0 Authorization Request as described in \sect{oauth2} (1-4) to obtain an \ac{at} for the \ac{ict} Request.
For this purpose, the \ac{at} requires a Scope sufficient to access the \ac{eu}'s profile information, e.g., \texttt{profile}, and an \ac{e2e} Scope, e.g., \texttt{e2e\_auth\_email}.
% Abruf von Identity Certification Token; Einführung Client + OP Key Pairs:
The Client then uses the \ac{at} to authorize an OAuth 2.0 Resource Request for an \ac{ict} (5) from the \ac{op}, called an \ac{ict} Request.
For this purpose, the Client uniquely generates a new public key $K^+_C$ and presents it to the \ac{op}.
The Client also presents a \ac{pop} of the corresponding private key $K^-_C$, e.g., by signing a unique nonce.
The \ac{op} verifies the validity of the \ac{at} and the \ac{pop} (6).
If valid, the \ac{op} signs the \ac{ict} with its private key $K^-_{\ac{op}}$ corresponding to its published public key $K^+_{\ac{op}}$ and responds with the \ac{ict} (7).
When the \ac{ict} expires and a new \ac{ict} is required, the Client repeats steps (5) to (7) to request a new \ac{ict} for a new key pair.

% \vspace{-0.2cm}
\subsection{\ac{e2e} Authentication with \ac{ict}}
\label{sec:ict_usage}
% \vspace{-0.1cm}
%-----------------------------------
% Einführung; Verweis auf Grafik:
The Client uses the \ac{ict} to authenticate its \ac{eu} to the \ac{ap}'s \ac{au} as shown in \fig{oidc2_e2ea}.
% Transfer von ICT; Erklärung PoP Möglichkeiten:
First, the Client passes the \ac{ict} containing its public key $K^+_C$ to the \ac{ap} and provides a \ac{pop} for the corresponding private key $K^-_C$ (1).
To do this, the Client signs either a unique nonce provided by the \ac{ap} or a unique session-specific identifier.
Alternatively, the Client can prove the possession by establishing a secure channel based on the private key $K^-_C$.
In \sect{applications}, we show and explain use cases that take advantage of these three options.
% Verifikation von ICT und PoP:
The \ac{ap} then verifies the Client's \ac{pop} (2) using the public key $K^+_C$ from the \ac{ict} and verifies the \ac{au}'s trust relationship with the \ac{op} (3).
This may require user interaction or the use of whitelists, discussed further in \sect{trust_relationship}.
If the \ac{au} trusts the \ac{op}, the \ac{ap} checks the expiration date and verifies the signature of the \ac{ict} using the \ac{op}'s public key $K^+_{\ac{op}}$ (4).
If successful, the \ac{eu} has proven its \ac{sso} identity to the \ac{au} (5).

%-------------------------------------------------------------------------------
% \vspace{-0.2cm}
\section{Security Considerations}
\label{sec:security}
% \vspace{-0.2cm}
%-------------------------------------------------------------------------------
% Überblick über Section:
First, we discuss how \ac{oidc²} shifts the burden of thorough authentication from service providers to identity providers.
Then, we analyze the trust relationship between \ac{oidc²} entities and propose a trust classification for \acp{op}.
Finally, we propose authentication with multiple \acp{ict} and discuss the correlation between the validity of an \ac{ict} and its corresponding key pair.

% \vspace{-0.2cm}
\subsection{Service Provider vs. OpenID Provider}
\label{sec:sp_vs_op}
% \vspace{-0.2cm}
%-----------------------------------
% Vorteile für Nutzer (e2e authentication); Vorteile für service provider (kaum Implementierungsaufwand, kein OP Betrieb nötig):
In most communication services, users must rely on the identity claims of their communication partners provided by the service provider, with no way to verify them.
\ac{oidc²} allows users to verify each other's identities without having to trust the service provider.
This only requires the Client to implement \ac{oidc²} and the protocol to provide a way to exchange the \acp{ict}.
The service provider does not need to implement OAuth 2.0 for the Client or provide an \ac{op}.
This improves the overall security of the service and prevents privacy issues by eliminating the need for the service provider to collect sensitive information about its users.

\subsection{Trust Relationship}
\label{sec:oidc2_trust_relationship}
\begin{figure}[ht]
    % \vspace{-0.8cm}
    \centering
    \includegraphics[width=\linewidth]{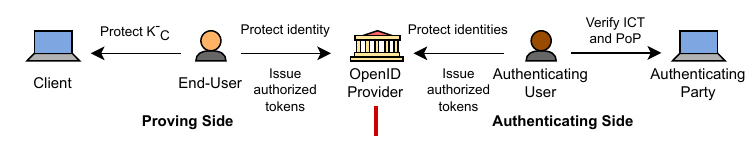}
    \caption{Trust relationship between OIDC²'s entities.}
    \label{fig:oidc2_trust_relationship}
    % \vspace{-0.6cm}
\end{figure}

% \vspace{-0.2cm}
%-----------------------------------
% Verweis auf Abbildung:
\fig{oidc2_trust_relationship} shows an overview of the trust relationship between the entities of the \ac{oidc²} protocol.

% Vertrauen auf Proving Side: EU -> OP; EU -> Client:
On the proving side, the \ac{eu} trusts his \ac{op} to protect his identity from impersonation attacks and not to impersonate him.
This includes that the \ac{op} will only issue authorized \acp{ict}.
Furthermore, the \ac{eu} trusts that his Client will operate as intended.
This means that the Client will protect its private key $K^-_C$ from third parties and use the \ac{ict} only for the intended authentication processes.
To limit potential misuse by the Client, the \ac{ict} is scoped to a specific context.
For example, this prevents an email client from misusing the \ac{ict} to sign contracts on behalf of the \ac{eu}.

% Vertrauen auf Authenticating Side: AU -> OP; AU -> AP:
On the authentication side, the \ac{au} trusts the \ac{op} to protect the \ac{eu}'s identity and to sufficiently verify the Client's possession of its private key $K^-_C$.
The \ac{au} also trusts the \ac{op} to certify sufficiently trustworthy identity claims with the issued \ac{ict}, which we will discuss in more detail in \sect{classification}.
To ensure that the authentication process is intended by the \ac{eu}, the \ac{au} trusts the \ac{op} to issue only \ac{eu}-authorized \acp{ict}.
The \ac{au} must also trust his \ac{ap} to correctly verify the received \ac{ict} and \ac{pop}.

% Lösungsansätze, wie Vertrauen zwischen AU und OP hergestellt werden kann:
The \ac{au} needs to trust the \ac{op}.
We offer two solutions that can be combined.
First, the \ac{ap} trusts a trusted identity federation such as the \ac{gain} \cite{GainWhitepaper}, which consists of international \acp{op} such as banks, insurance companies, or government institutions, all of which manage fully verified real-world identities.
Second, the \ac{au} maintains his own whitelist of \acp{op}, such as social media platforms or his business partners.
Not every \ac{op} has the same level of trustworthiness, so we classify them in the next section.

% \vspace{-0.4cm}
\subsection{Classification of OpenID Providers}
\label{sec:classification}
% \vspace{-0.1cm}
%-----------------------------------
When working with \ac{oidc²}, we suggest three classes of \acp{op} to consider.

% \vspace{-0.4cm}
\subsubsection{\jp{\ac{iop}}}
\label{sec:insecure_op}
% Beschreibung von nicht vertrauenswürdigem OP:
\acp{op} can be considered insecure for a variety of reasons.
They may not be able to sufficiently protect their users' credentials, or they may be untrustworthy for political or economic reasons.
For example, they may certify potentially false or insufficiently verified claims.
If an \ac{au} considers an \ac{op} insecure, his \ac{ap} will not accept any \acp{ict} issued by that \ac{op}.

% \vspace{-0.4cm}
\subsubsection{\jp{\ac{aop}}}
\label{sec:authoritative_op}
% AOP schützt accounts hinreichend und ist Autorität für bestimmte claims:
We classify an \ac{op} as an \ac{aop} for specific claims, if the \ac{au} accepts the \ac{op} as an authority for those claims and trusts the \ac{op} to protect managed identities.
For example, an email server's \ac{op} is authoritative for email addresses within its own domain.
Because an \ac{op} issues a unique subject identifier for each \ac{sso} identity by specification, an \ac{op} is always authoritative for its associated \texttt{sub} claim.

% Beispiel und Limitations:
This makes \acp{aop} sufficient for scenarios where an \ac{eu} wants to be authenticated with specific claims.
For example, if the \ac{au} knows the \ac{eu}'s email address, the \ac{eu} uses an \ac{ict} issued by his email provider's \ac{op} to authenticate on a social media platform.
However, \acp{aop} are only sufficient to certify identity claims for which they are an authority.
To certify real-world identity claims such as names or addresses, the \ac{aop} must typically be the \ac{op} of a trusted government organization.

% \vspace{-0.4cm}
\subsubsection{\jp{\ac{vop}}}
\label{sec:verifying_op}
% Ein VOP schützt seine Accounts hinreichen und verifiziert bestimmte claims hinreichend:
There is not always an \ac{aop} for every claim the \ac{eu} wants to be authenticated with.
Instead, the \ac{eu} can use a third-party service that the \ac{au} trusts to sufficiently verify his identity claims and protect his account.
We call the \ac{op} of this third-party service a \jp{\ac{vop}}.
This \ac{vop} could check the \ac{eu}'s passport to verify his name, or send him a verification code via SMS to verify his phone number.

% Beispiele, Limitations und Hybrid-Modelle:
There are already \jp{\acs{op}} such as banks or insurance companies that are required by law to verify their customers' claims.
However, such verification processes are often costly, which is why \acp{vop} often do not verify all claims or offer it as an optional service, such as the social media platforms Facebook and \jp{X}.
Both can be \acp{aop} at the same time.
For example, banks are \acp{vop} for the name of an \ac{eu}, but also \acp{aop} for bank account numbers.

% \vspace{-0.3cm}
\subsection{Authentication with Multiple \acp{ict}}
\label{sec:multi_ict}
% \vspace{-0.1cm}
%-----------------------------------
% Diskussion über Vor- und Nachteile der Authentifizierung mit mehreren ICTs.
The classification of an \ac{op} is up to the \ac{au}, i.e., the \ac{au} may not accept \acp{ict} from certain \acp{op}.
Since an \ac{eu} may not know the \ac{au}'s classification in advance, the \ac{eu} can present \acp{ict} from different \acp{op} and the corresponding \acp{pop} to increase the likelihood of successful authentication by the \ac{au}.
However, this requires more work for the \ac{eu} as he has to log in to all these \acp{op} to receive \acp{ict}.
If the \ac{ap} receives multiple \acp{ict}, it presents them to the \ac{au}, which then selects the most trusted issuer or rejects them all.
Furthermore, the \ac{eu} must be aware that presenting multiple \acp{ict} also exposes all his presented accounts to the \ac{au}.

% \vspace{-0.3cm}
\subsection{Validity of \acp{ict} and Client Key Pairs}
\label{sec:ict_validity_period}
% \vspace{-0.1cm}
%-----------------------------------
% Einführung, wie das ICT mit dem Key Pair des Clients zusammenhängt:
An \ac{ict} contains the public key $K^+_C$ of the Client.
By issuing this \ac{ict}, the ac{op} certifies that the corresponding \ac{eu} authorized the Client for \ac{e2e} authentication with the contained identity claims.
% By issuing an ICT, the OP certifies that the Client whose public key $K^+_C$ the ICT contains is authorized by the EU to perform e2e authentication with the contained identity claims.
An attacker trying to impersonate the \ac{eu} needs the corresponding private key of the \ac{ict}.
% Whoever proves to be in possession of the corresponding verified private key $K^-_C$ is also authorized -- even attackers.
% Einschränkung von ICT durch validity period und scope:
We minimize this potential misuse of the \ac{ict} by a leaked private key by making the \ac{ict} short-lived and limited in scope.
Since a few minutes are sufficient for most use cases (see \sect{applications}), we recommend setting the \ac{ict} validity period to no more than 1 hour.

% ICT Validity = Key Pair Validity:
We propose that an ephemeral and unique key pair $K^\pm_C$ expires along with its associated \ac{ict}, eliminating the need for key revocation mechanisms.
% Ausnahme ist long-term key pair; Erfordert Verweis auf key revocation list:
However, Sections \ref{sec:im} and \ref{sec:email} show that long-term key pairs are useful in some cases.
Therefore, we further propose that an \ac{ict} may also contain a long-term public key, which must be indicated by providing the key revocation server of the key.
Such a key is valid until revoked and is associated with the claims in the \ac{ict}.
Some services control the lifetime of public keys by associating them with user profiles.
An example of this approach is the Signal protocol (see \sect{im}).
In such applications, a user can be authenticated with a public key received from an \ac{ict} as long as the public key contained in it is associated with the profile.
% Session darf auch nach key expiry gültig bleiben:
In any case, an active session can remain valid even after the underlying key pair $K^\pm_C$ expires (see \sect{vc}).

%-------------------------------------------------------------------------------
% \vspace{-0.2cm}
\section{Use of \ac{oidc²} in Applications}
\label{sec:applications}
% \vspace{-0.2cm}
%-------------------------------------------------------------------------------
% Überblick über diese Section:
\jp{We explain how \ac{e2e} authentication is currently implemented in video conferencing, \ac{im}\jp{,} and email, and how it can be improved by \ac{oidc²}.}
\jp{The use of \ac{oidc²} is not limited to these applications, they were just selected to illustrate different application patterns.}
In addition, we recommend validity periods for \acp{ict} depending on these applications.

% \vspace{-0.2cm}
\subsection{Video Conferencing}
\label{sec:vc}
% \vspace{-0.1cm}
%-----------------------------------
% Überblick über die Section:
\jp{Most video conferencing systems do not utilize any \ac{e2e} authentication.}
\jp{Instead,} users must rely on the identities of their communication partners provided by the service's \ac{idp}.
\jp{Identifying a communication partner just by its video does not suffice.}
\jp{New technologies like deep fakes make this unreliable as demonstrated by an incident in 2022 \cite{Oltermann2022}.}
We explain how video conferencing services use OAuth 2.0 and \ac{oidc} and how they can benefit from \ac{oidc²}.

% \vspace{-0.2cm}
\subsubsection{\jp{\ac{e2e}} Authentication in Video Conferencing}
% Einführung wie Authentifizierung in VCs mit OAuth 2.0 und OIDC funktioniert:
In video conferencing, users log in to the service provider's OAuth 2.0 \ac{as} either directly with their credentials\jp{,} or through the \ac{op} with their \ac{oidc} identity, as explained in \sect{sso}.
After authentication, the video conferencing service provider's \ac{as} \jp{obtains} an \ac{idt} from the \ac{op}.
After authorization, the Client gets an \ac{at} from the \ac{as}.
The Client \jp{proves} its authorization to the video conferencing server \jp{with the \ac{at}}.
\jp{The video conferencing server retrieves the \ac{eu}'s service account ID from the \ac{at} and provides the corresponding service profile to the communication partner.}
\jp{Note that the \ac{eu}'s service profile may be different from his \ac{oidc} identity.}

% Einführung in die E2EE und Vertrauensproblematik:
\jp{In addition, the clients of both communication partners generate an ephemeral asymmetric key pair.}
\jp{They sign their public keys and key negotiation messages and exchange them via the video conferencing server.}
\jp{This enables an \ac{e2e} encrypted communication channel, but users cannot rely on their communication partner’s identity for two reasons.}
\jp{First, the service profile may not reflect the partner's real-world identity.}
\jp{Second, the service provider may provide a fake profile.}

% \vspace{-0.2cm}
\subsubsection{\jp{\ac{e2e}} Authentication with \ac{oidc²}}
\begin{figure}[ht]
    % \vspace{-0.5cm}
    \centering
    \includegraphics[width=\linewidth]{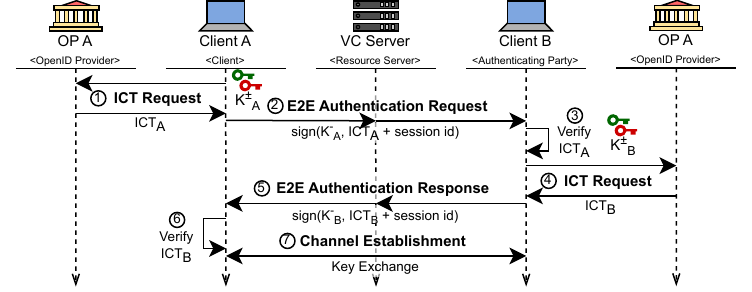}
    % \vspace{-0.7cm}
    \caption{\ac{e2e} authentication with \ac{oidc²} for video conferencing.}
    \label{fig:oidc2_vc}
    % \vspace{-0.6cm}
\end{figure}

% Grobe übersicht; Verweis auf Grafik:
We propose that the \ac{eu} \jp{is authenticated by his communication partner (\ac{au}) with his \ac{ict} that the \ac{eu}'s Client requested from the \ac{op}}.
After a mutual \ac{ict} exchange, the Client and the \jp{\ac{ap}} use the contained verified public keys to establish a secure channel, as shown in \fig{oidc2_vc}.

% Verbindungsanfrage Client -> AP:
First, Client A generates an ephemeral key pair $K^\pm_A$ and \jp{requests an \ac{ict} for its public key $K^+_A$ from the \ac{eu}'s \ac{op}} (1).
\jp{The Client signs this \ac{ict} and some unique session context, e.g., including timestamps and ephemeral public keys, with its private key $K^-_A$.}
\jp{The latter serves as a \ac{pop}.}
\jp{The Client sends the \ac{ict} and the \ac{pop}} to the \ac{ap} via the video conferencing server (2).
% Verbindungsannahme AP -> Client:
If the \ac{au} trusts the \ac{eu}'s \ac{op}, \jp{it validates the \ac{ict} and verifies the \ac{pop}.}
\jp{Then} Client B generates its own ephemeral key pair $K^\pm_B$ (3) and requests an \ac{ict} \jp{for its public key $K^+_B$} from the \ac{au}'s \ac{op} (4).
The \ac{ap} signs its \ac{ict} and the \jp{unique session context with its private key $K^-_B$ (\ac{pop})} and responds to \jp{Client A} via the video conferencing server (5).
% Verbindungsaufbau:
If the \ac{eu} trusts the \ac{au}'s \ac{op} (\jp{6}) \jp{an the validation of the \ac{ict} and the \ac{pop} are successful}, then the Client and \jp{the} \ac{ap} have successfully performed mutual authentication\jp{, which enables} them to establish a \jp{securely \ac{e2e} authenticated and encrypted channel} (7).

% \vspace{-0.4cm}
\subsubsection{Discussion}
% Größter Vorteil von OIDC².
\jp{When video conferencing is improved with \ac{oidc²}, users do no longer need to trust their service provider to display the correct service profiles of the participants in a video session.}
\jp{While the service profiles may not reflect the true identities of the participants, the users can rely that verifying \acp{op} have thoroughly checked the participants' identity when they registered.}
% Validity Period of ICT:
\jp{We recommend a validity period of 5 minutes for video conferencing services as starting} a video conference takes only a few seconds.
\jp{Within that time a user has joined a call, and that duration is also long enough to compensate for clock drifts on end systems.}
When the \ac{ict} expires, an active secure channel remains valid until it is closed.

% \vspace{-0.3cm}
\subsection{\jp{\acf{im}}}
\label{sec:im}
% \vspace{-0.1cm}
%-----------------------------------
% Section-Übersicht:
\jp{Secure \ac{e2e} authentication in \ac{im} requires a prior secure exchange of public keys which is mostly achieved by face-to-face meetings.}
%\jp{We suggest how the Signal \cite{signal} \ac{im} service could benefit from \ac{oidc²}, eliminating the need for face-to-face meetings.}
\jp{We suggest an \ac{oidc²}-based authentication method for \ac{im} which does not require such a prior secure exchange of public keys.}

% \vspace{-0.4cm}
\subsubsection{\jp{\ac{e2e}} Authentication in Signal}
% Einführung in Ende-zu-Ende Authentifizierung in Signal und In-Person Verifikation mit QR Codes:
In the Signal protocol \cite{WhatsappWhitepaper}, users are identified by their phone number and public key $K^+$\jp{.}
\jp{Both phone number and public key $K^+$ are verified and published by Signal while the corresponding long-term private key $K^-$ remains on the user's device.}
\jp{To establish an \ac{e2e} encrypted communication channel, a user requests from Signal the public key of the communication partner.}
\jp{The public keys of both users are leveraged for an authenticated Diffie-Hellman key exchange to establish an encrypted communication channel between the partner.}
\jp{Over this channel, they may authenticate each other with their public keys.}
\jp{However, at this point the public keys are not yet reliable as this method requires trust in Signal and its verification method for phone numbers and public keys.}
\jp{When the partners meet in presence, they can mutually verify each other's public keys by exchanging them via a secure side channel, e.g., by presenting them as a \ac{qr} code in a face-to-face meeting.}
%\jp{This manual process eliminates the need for trust in Signal.} 
\jp{When the partners communicate again, they can rely on the verified public keys and trust in Signal is no longer needed.}
\jp{This mechanism is a strong but also cumbersome.}

% \vspace{-0.4cm}
\subsubsection{\jp{\ac{e2e}} Authentication with \ac{oidc²}}
\begin{figure}[ht]
    % \vspace{-0.5cm}
    \centering
    \includegraphics[width=\linewidth]{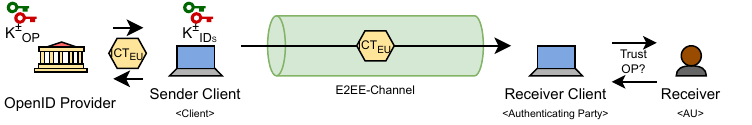}
    % \vspace{-0.4cm}
    \caption{\jp{Unilateral \ac{e2e} \ac{im} authentication with \ac{oidc²}.}}
    \label{fig:e2e_im_authentication}
    % \vspace{-0.8cm}
\end{figure}

% Einführung; Verweis auf Figure.
We propose an \ac{e2e} authentication method for \ac{im} based on \ac{oidc²}.
\jp{It is} illustrated in \fig{e2e_im_authentication}.
% Kurze Erklärung, wie Kanal mit OIDC² nachträglich authentifiziert wird:
\jp{We assume that the \ac{im} clients have already established an \ac{e2e} encrypted channel that is mutually authenticated by each other's public keys $K'^+$.}
\jp{The mutual authentication method based on \ac{oidc²} works as follows.}
\jp{The} \ac{im} \jp{client} requests an \ac{ict} from its \ac{eu}'s \ac{op} for his public key $K^+$ and sends the \ac{ict} over the secure channel to the \ac{ap}.
If the \ac{au} trusts the \ac{eu}'s \ac{op}, the \ac{ap} verifies the received \ac{ict} and compares the contained public key $K^+$ with the \jp{one that was used to establish the secure channel.}
\jp{The implicit \ac{pop} consists of the fact that both users can communicate over the secure channel, i.e., they possess the corresponding private keys.}

% \vspace{-0.4cm}
\subsubsection{Discussion}
\jp{The presented approach shows that existing key management systems like the one of Signal can be extended with \ac{oidc²} as an authentication layer.}
\jp{It} does not use any Signal-specific features and can therefore be applied to any other \ac{im} service \jp{while preserving all security-related features such as ratchet-based \ac{e2e} encryption or forward secrecy.}
\jp{As a particularity, the \ac{ict}'s key is not ephemeral but an existing long-term key. Moreover, the method demonstrates that a secure channel may be set up with non-verified public keys with subsequent key verification based on \ac{oidc²}. Probably, this pattern can also be applied to other applications.}
% Validity Time:
\jp{We recommend a validity period of 5 minutes for \acp{ict} in an \ac{im} context because messages are delivered to the receiving client very quickly.}
\jp{If the \ac{ict} is transmitted when the \ac{ap} is offline, the verification process must be repeated.}

\subsection{Email}
\label{sec:email}
% \vspace{-0.2cm}
%-----------------------------------
% Übersicht über Section:
\jp{\ac{smime} and \ac{pgp} are common standards for email authentication for almost three decades.}
\jp{However, probably due to their complex key management, signed emails are still the exception \cite{Stransky2022} with $2.8 \%$.}
\jp{We briefly describe \ac{smime} and \ac{pgp} and their shortcomings, explain how \ac{pgp} can be enhanced with \ac{oidc²} for better authentication, and the limitations of that approach.}

% \vspace{-0.5cm}
\subsubsection{\jp{\ac{e2e} Authentication with \ac{smime} and \ac{pgp}}}
% Einführung Email-Signaturen:
\jp{\ac{smime} and \ac{pgp} utilize a long-term asymmetric key pair $K^\pm$ to sign emails.}
% Einführung S/MIME:
\jp{\ac{smime} leverages X.509 certificates issued by a \ac{ca} so that receivers of a signed email can validate its signature with the enclosed public key after validation of the public key and checking its associated \ac{crl}.}
\jp{Obtaining such a certificate may be a cumbersome and expensive process for the user unless it is provided by his employer.}
% Einführung PGP:
\jp{\ac{pgp} requires that the receiver of an email has obtained a fingerprint of the sender's public key via some side channel.}
\jp{Mostly, the communication partners have exchanged the fingerprints of their public keys in a face-to-face meeting.}
\jp{This is also a cumbersome process and does not allow for signed communication before having known the receiver of the email.}
\jp{To facilitate revocation of a key pair, the public key is published on a key server.}
\jp{The key server maintains a key revocation list which needs to be checked by the receiver of an email.}
% Thus, the receiver of an email must have met with the sender before, identify the sender by the received public key, and check the \ac{crl} on the corresponding key server.

% Einführung Email Verschlüsselung:
\jp{\ac{smime} and \ac{pgp} also support email encryption.}
\jp{To that end, the sender encrypts an email with a symmetric key, encrypts this key with the receiver's public key $K^+$, and attaches it to the email.}
\jp{The receiver uses his private key $K^-$ to decrypt the symmetric key for decryption of the email.}

% Zusammenfassung der Probleme:
\jp{Both \ac{smime} and \ac{pgp} suffer from the fact that they require complex key management.}
\jp{First, the user must protect his private key $K^-$.}
\jp{Second, the user must securely synchronize his key pair $K^\pm$ across his devices if he wants to use them all for signed and encrypted email communication.}
\jp{And third, the sender must revoke compromised key pairs $K^\pm$, and the receiver must check \acp{crl} to validate the validity of a received public key $K^+$.}

% \vspace{-0.4cm}
\subsubsection{\jp{\ac{e2e} Authentication with \jp{\ac{pgp} and} \ac{oidc²}}}
% Vorstellung von PGP + OIDC²:
\jp{We propose to combine \ac{pgp} with the key verification method of \ac{oidc²}, but do not touch any security properties of \ac{pgp}.}
\jp{We have implemented this concept in a prototype which is published on GitHub\footnote{\url{https://anonymous.4open.science/r/oidc2-demo/}}.}

\begin{figure}[ht]
    % \vspace{-0.6cm}
    \centering
    \includegraphics[width=\linewidth]{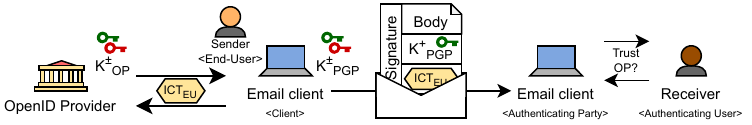}
    \caption{\jp{\ac{e2e} email authentication with \ac{pgp} and \ac{oidc²}.}}
    \label{fig:e2e_email_authentication}
    % \vspace{-0.6cm}
\end{figure}

% Signieren und Versenden der Email:
\jp{\fig{e2e_email_authentication} shows how emails are sent with \ac{pgp} and \ac{oidc²}.}
\jp{The sender (\ac{eu}) authorizes his email client (Client) for \ac{e2e} authentication in the \texttt{email} context.}
\jp{Then the Client requests an $\ac{ict}_{\ac{eu}}$ for its stored \ac{pgp} key $K^\pm_{\ac{pgp}}$, attaches both the public \ac{pgp} key $K^+_{\ac{pgp}}$ and the obtained $\ac{ict}_{\ac{eu}}$ to the email, and signs the email with the private \ac{pgp} key $K^-_{\ac{pgp}}$.}
% Empfangen und verifizieren der Email:
\jp{The receiver (\ac{au}) opens the email with his client (\ac{ap}).}
\jp{The \ac{ap} first validates the $\ac{ict}_{\ac{eu}}$, i.e., it checks whether it trusts the issuing \ac{op} of the \ac{ict} and it validates the \ac{ict}'s signature.}
\jp{This can be done offline as the public keys of trusted \acp{op} are typically cached for a moderate time.}
\jp{Then, the \ac{ap} verifies the email's integrity by verifying its signature with the \ac{ict}'s contained public key.}
% Erklärung von PoP:
\jp{The integrity of the mail serves as a \ac{pop} to prove that the sender of the email is also the owner of the key in the $\ac{ict}_{\ac{eu}}$.}
\jp{Now, the sender of the email is identified with the identity provided in the $\ac{ict}_{\ac{eu}}$.}

% \vspace{-0.5cm}
\subsubsection{Discussion}
% Vorteil von OIDC²: Kein Secure Side Channel mehr notwendig
\jp{The email is sufficiently authenticated when it was validated before the expiration of the $\ac{ict}_{\ac{eu}}$.}
\jp{Then, checking a possibly available \ac{crl} of the sender's PGP key is not needed, as the validity of the key is already given by the $\ac{ict}_{\ac{eu}}$.}
\jp{When receiving signed emails with conventional \ac{pgp} or \ac{smime}, the key's \ac{crl} must be checked so that the \ac{crl} must be reachable.}
\jp{Conversely, \ac{oidc²}-based \ac{pgp} requires that the \ac{op} is reachable to issue the $\ac{ict}_{\ac{eu}}$ when sending the email.}

\jp{The \ac{pgp} key pair in the presented method can be either the sender's long-term \ac{pgp} key pair, or a short-term key pair just generated by the Client.}
% Diskussion zu long-term key pair:
\jp{We first assume that a long-term public \ac{pgp} key is attached to the email and the contained $\ac{ict}_{\ac{eu}}$ references a key server that is reliable in the sense that it adds information reliably to the \ac{crl}.}
\jp{The information about the public key on the key server equals the one in the $\ac{ict}_{\ac{eu}}$ and that this information predates the issuing time of the $\ac{ict}_{\ac{eu}}$.}
\jp{These preconditions seem comprehensive but are mostly fulfilled.}
\jp{We can argue that the \ac{op} confirmed the mapping of key and identity, which was equal to the information on the key server at that time, so that this information on the key server is also reliable as long as the key is valid, i.e., until it expires or it is revoked.}
\jp{Therefore we can retain two advantages of long-term keys as long as the key is valid.}
\jp{First, the message may be even validated after the $\ac{ict}_{\ac{eu}}$ expired.}
\jp{Second, the key may be stored as an authenticated public \ac{pgp} key of the sender.}
\jp{Thus, the proposed procedure may be used to securely exchange a long-term \ac{pgp} key, and this key may be used to send encrypted emails to the key's owner.}

% Diskussion zu short-term key pair:
\jp{We now consider the use of short-lived keys.}
\jp{They do not need to be stored so that no complex key management is required.}
\jp{This is unlike for long-term keys and greatly improves the usability of email authentication.}
\jp{However, short-lived keys come with two drawbacks.}
\jp{First, sending encrypted emails is not possible without long-term keys.}
\jp{Second, an email can be successfully verified only as long as the $\ac{ict}_{\ac{eu}}$ is valid, which may be too short when the email is received late by the \ac{ap}.}
\jp{This problem can be solved if the \ac{eu} can trust his mail server so that the \ac{ap} can rely on the inbox timestamp.}
\jp{Then an email can be considered valid if it was received within the validity period of the attached \ac{ict}.}
\jp{As most emails are received within a few minutes by the inbox on the mail server, we recommend an expiration date of at most one hour.}

%-------------------------------------------------------------------------------
% \vspace{-0.4cm}
\section{Implementation}
\label{sec:implementation}
% \vspace{-0.2cm}
%-------------------------------------------------------------------------------
We present a simple extension for any \ac{oidc} server to handle \ac{ict} Requests including a \ac{pop} for the verification of the Client's public key.
The implementation is available on GitHub\footnote{\url{https://github.com/oidc2/op-extension}}.
However, we recommend it only for testing purposes.

To request a token, a Client sends a Token Request to the so-called \texttt{/token} Endpoint of the \ac{op}.
That is a special path in the URL of the \ac{oidc} server.
Moreover, there is also a \texttt{/userinfo} Endpoint that returns information about the user upon a Userinfo Request.

Many services are not directly reachable on the Internet but via a reverse proxy.
A reverse proxy is an \ac{http} server that resides in the same network as the server, terminates the TLS connection between client and server, and relays data to and from the application server from and to the client.

\begin{figure}[ht]
    % \vspace{-0.4cm}
    \centering
    \includegraphics[width=.7\linewidth]{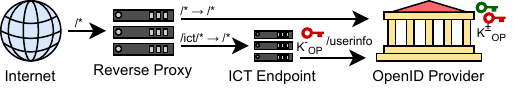}
    \caption{Simple extension to an \ac{oidc} server to handle \ac{ict} Requests.}
    \label{fig:oidc2_implementation_architecture}
    % \vspace{-0.6cm}
\end{figure}

We propose the generic extension to an \ac{oidc} server in \fig{oidc2_implementation_architecture} so that the \ac{oidc} server can handle \ac{ict} Requests.
We define a novel \texttt{/ict} Endpoint which runs as a microservice separately from the \ac{oidc} server.
The \texttt{/ict} Endpoint and the \ac{oidc} server operate behind a reverse proxy.
The reverse proxy forwards any conventional \ac{oidc} requests to the \ac{oidc} server and \ac{ict} Requests to the \texttt{/ict} Endpoint.

The \texttt{/ict} Endpoint expects an \ac{at} with Scopes for identity claims, e.g., \texttt{profile} for name and birth date, and a scoped context for \ac{e2e} authentication, e.g., \texttt{e2e\_auth\_email} for the email context, in the \ac{ict} Request.
It extracts the \ac{at}, and includes it in a Userinfo Request to the \ac{oidc} server.
After \jp{receiving} user information, the \texttt{/ict} Endpoint checks whether the \ac{eu} possesses the private key $K^-_C$ for the public key $K^+_C$ contained in the \ac{ict} request, which is explained later.
If the check was successful, the \texttt{/ict} Endpoint issues an \ac{ict} with appropriate information and signs it with the private key $K^-_{\ac{op}}$ of the \ac{op}.
Thus, $K^-_{\ac{op}}$ must be available to the \texttt{/ict} Endpoint.
This is a reason why we recommend this simple prototype only for testing purposes but not for production.
Finally, the \texttt{/ict} Endpoint returns the \ac{ict} to the Client.

To save communication overhead between the \texttt{/ict} Endpoint and the Client, we propose the following \ac{pop}.
The Client chooses a nonce, concatenates it with a timestamp, signs the concatenation with its private key $K^-_C$, and includes concatenation and signature in the \ac{ict} Request.
The \texttt{/ict} Endpoint verifies the signature with the public key $K^+_C$ and caches the nonce for 30 seconds.
To counter replay attacks, the \texttt{/ict} Endpoint accepts only \ac{ict} Requests with timestamps in the concatenation that deviate at most 15 seconds from its own time and whose nonce is not in the cache.

%-------------------------------------------------------------------------------
% \vspace{-0.2cm}
\section{Evaluation}
\label{sec:evaluation}
% \vspace{-0.2cm}
%-------------------------------------------------------------------------------
We evaluate the performance of the provided \texttt{/ict} Endpoint, written in Go, compared to the \texttt{/token} Endpoint of the Keycloak~22.0.1 and Authentik~2023.6.1 \ac{oidc} server software \jp{to estimate the additional costs for an \ac{op}}.
% They are written in Java and Go, respectively.

We conduct the following two experiments.
(A) A Client sends a Refresh Token to the \texttt{/token} Endpoint of the \ac{oidc} server and obtains an \ac{idt}, an \ac{rt}, and an \ac{at}.
(B) A Client generates a \ac{pop}, sends an \ac{at} to the new \texttt{/ict} Endpoint, and obtains an \ac{ict}.
Both experiments are conducted over one minute, i.e., a token is requested, returned, and then the next request is sent.
We ran each experiment 20 times and computed mean requests per minute including confidence intervals with a confidence level of 95\% (${CI}_{0.95}$) using the Student's t-distribution.
We automate this process with the help of a web application\footnote{The application is programmed in Angular~15 and its code is available on GitHub \url{https://github.com/oidc2/benchmark}}. 

The \ac{oidc} server, its user database based on PostgreSQL~15.2, and the new \texttt{/ict} Endpoint run in separate Docker containers\footnote{\url{https://github.com/oidc2/op-extension/blob/main/docker-compose.yaml}}.
The host is a Lenovo ThinkPad T14s with \jp{a} 2.1 GHz AMD Ryzen 5 PRO 4650U processor, 16 GB RAM, and a 512 GB SSD with Windows~11 22H2 x64, and running the Docker engine\footnote{\url{https://www.docker.com/}}~24.0.2 in WSL~2\footnote{\url{https://learn.microsoft.com/en-us/windows/wsl/}}.
While Authentik can import and export any private keys, Keycloak cannot export private keys and it can import only RSA keys.
Therefore, we chose RS256 for signatures, i.e., a 2048 bit RSA key with the SHA-256 hashing algorithm to make experiments with different server software comparable.

With Keycloak, a mean request rate of 994.00 \acp{idt}\jp{/min} (A) (${CI}_{0.95}$: [992.97; 995.03]) and 988.20 \acp{ict}\jp{/min} (B) (${CI}_{0.95}$: [986.72; 989.68]) could be served\footnote{The values per experiment run are available here: \url{https://github.com/oidc2/benchmark/blob/main/results}.}.
In contrast, with Authentik, 190.95 \acp{idt}\jp{/min} (A) (${CI}_{0.95}$: [190.35; 191.35]) and 891.65 \acp{ict}\jp{/min} (B) (${CI}_{0.95}$: [886.04; 897.26]) could be served.
Thus, the tested version of Keycloak is more efficient than the tested version of Authentik.
Moreover, the provided \texttt{/ict} Endpoint is as efficient as the built-in \texttt{/token} Endpoint or even more efficient.

We compare the work done by the \texttt{/token} Endpoint and the \texttt{/ict} Endpoint. 
(A) The \texttt{/token} Endpoint validates the \ac{rt}, creates an \ac{idt}, and signs the \ac{at} and the \ac{idt} with its private key.
The integrity of the \ac{rt} is secured differently\footnote{Authentik uses a nonce for the \ac{rt} stored in the database while Keycloak secures the \ac{rt} with an HMAC.}.
(B) The \texttt{/ict} Endpoint validates the \ac{pop} for the Client's public key, and requests user information using an \ac{at} from the \texttt{/userinfo} Endpoint, which validates the \ac{at}.
Then the \texttt{/ict} Endpoint creates and signs the \ac{ict}.

The effort for creating and signing an \ac{idt} in (A) and an \ac{ict} in (B) is possibly similar, as both require \ac{rt}/\ac{at} validation, a database request, and a token signature.
Thus, creating an \ac{rt} and \ac{at}, and signing the \ac{at} in (A) is apparently equal or more time consuming than creating the \ac{pop} at the Client and validating the \ac{pop} at the \texttt{/ict} Endpoint in (B).

\jp{The cost of providing \acp{ict} scales with the frequency \acp{ict} are requested, which depends on the adoption of \ac{oidc²} by applications and by \acp{eu}.}
\jp{However, \acp{ict} are typically requested by \acp{eu}' Clients before sending an email, when being authenticated by a new \ac{im} communication partner, and when joining a video conferencing session.}
\jp{Such user-triggered actions may not exceed 10 \ac{ict} requests per hour in normal workloads.}
\jp{Inactive \acp{eu} do not request any \acp{ict}.}
\jp{In contrast, \acp{at} are recommended to be renewed every hour to comply with \rfc{6749}'s expiration time guidance.}
\jp{This is needed for every running Client, even if the \ac{eu} is inactive.}
\jp{Therefore, our predicted usage of \acp{ict} after full adoption is in a similar order of magnitude as the recommended need for \acp{at}.}

%-------------------------------------------------------------------------------
% \vspace{-0.2cm}
\section{Conclusion and Future Work}
\label{sec:conclusion}
% \vspace{-0.2cm}
%-------------------------------------------------------------------------------
This paper introduced \ac{oidc²}, which allows \acp{eu} to request a verifiable \ac{ict} from an \ac{op}.
An \ac{ict} contains claims about an \ac{eu} and a public key chosen by the \ac{eu}.
\acp{au} can authenticate \acp{eu} with an \ac{ict} if they trust his issuing \ac{op}.
We compared \ac{oidc²} to existing \ac{e2e} authentication methods and found that \ac{oidc²} is easier to use and improves security by eliminating the need for revocation lists.
We suggested how \ac{oidc²} can be implemented based on the \ac{oidc} framework.
We discussed security considerations for and general improvements with \ac{oidc²}: the trust relationship among its entities, a classification of \acp{op} and their utilization with \ac{oidc²}, authentication with multiple \acp{ict} to increase the likelihood of successful authentication, as well as appropriate (short) validity periods for \acp{ict}.
Furthermore, we proposed how \ac{oidc²} can be used for simple and user-friendly \ac{e2e} authentication for video conferencing, email, and \ac{im}.
Finally, we provided a simple, open-source extension for \ac{oidc} server software to support \ac{oidc²} for testing purposes.
We proved its compatibility with Authentik and Keycloak and the performance of the new \texttt{/ict} Endpoints is comparable to or better than the performance of the existing \texttt{/token} Endpoints.

To demonstrate the feasibility of \ac{oidc²} for \ac{e2e} authentication, we plan to integrate \ac{oidc²} for video conferencing based on the open WebRTC protocol, for \ac{im} based on the open Matrix protocol, and for email communication based on the \ac{pgp} standard.

\newpage

\section*{List of Abbreviations}

\begin{acronym}

\acro{oidc}[OIDC]{OpenID Connect}
\acro{oidc²}[OIDC²]{Open Identity Certification with OpenID Connect}
\acro{op}[OP]{OpenID Provider}
\acro{as}[AS]{Authorization Server}
\acro{eu}[EU]{End-User}
\acro{ict}[ICT]{Identity Certification Token}
\acro{pgp}[PGP]{Pretty Good Privacy}
\acro{smime}[S/MIME]{Secure / Multipurpose Internet Mail Extensions}
\acro{pki}[PKI]{Public Key Infrastructure}
\acro{ssi}[SSI]{Self-Sovereign Identity}
\acro{http}[HTTP]{Hypertext Transfer Protocol}
\acro{rs}[RS]{Resource Server}
\acro{ro}[RO]{Resource Owner}
\acro{at}[AT]{Access Token}
\acro{pr}[PR]{Protected Resource}
\acro{rt}[RT]{Refresh Token}
\acro{rp}[RP]{Relying Party}
\acro{idt}[IDT]{ID Token}
\acro{ca}[CA]{Certificate Authority}
\acro{idp}[IdP]{Identity Provider}
\acro{saml2}[SAML2]{Security Assertion Markup Language 2}
\acro{jwt}[JWT]{JSON Web Token}
\acro{did}[DID]{Decentralized Identifier}
\acro{issuer}[I]{Issuer}
\acro{holder}[H]{Holder}
\acro{verifier}[V]{Verifier}
\acro{vc}[VC]{Verifiable Credential}
\acro{vp}[VP]{Verifiable Presentation}
\acro{siopv2}[SIOPv2]{Self-Issued OpenID Provider Version 2}
\acro{openid4vci}[OpenID4VCI]{OpenID for Verifiable Credential Issuance}
\acro{w3c}[W3C]{World Wide Web Consortium}
\acro{openid4vp}[OpenID4VP]{OpenID for Verifiable Resentations}
\acro{hsm}[HSM]{Hardware Security Module}
\acro{tpm}[TPM]{Trusted Platform Module}
\acro{cic}[CIC]{Client Instance Claim}
\acro{pk}[PK]{PubKey}
\acro{ap}[AP]{Authenticating Party}
\acro{au}[AU]{Authenticating User}
\acro{e2e}[E2E]{End-to-End}
\acro{pop}[PoP]{Proof of Possession}
\acro{gain}[GAIN]{Global Assured Identity Network}
\acro{iop}[IOP]{Insecure OpenID Provider}
\acro{aop}[AOP]{Authoritative OpenID Provider}
\acro{vop}[VOP]{Verifying OpenID Provider}
\acro{im}[IM]{Instant Messaging}
\acro{sso}[SSO]{Single Sign-On}
\acro{crl}[CRL]{Certificate Revocation List}
\acro{qr}[QR]{Quick-Response}
% \acro{}[]{}

\end{acronym}

\bibliographystyle{unsrt}
\bibliography{literature}

\end{document}